\documentclass[12pt]{article}
\usepackage{graphicx,color}
\usepackage[polish,english]{babel}
\newcommand{\be}{\begin{equation}}
\newcommand{\ee}{\end{equation}}

\newcommand{\ba}{\hspace*{-5pt}\begin{array}}
\newcommand{\ea}{\end{array}}

\begin{document}
\sloppy

\vspace{6 mm}

\begin{center}
{\bf  EXACT SOLUTION OF THE HYPERBOLIC GENERALIZATION OF
BURGERS EQUATION, DESCRIBING TRAVELLING FRONTS AND THEIR INTERACTION }

\vspace{6mm}

{\Large Vsevolod Vladimirov }

\vspace{6mm}

{\it
AGH University of Science and Technology\\
Faculty of Applied Mathematics \\
al. Mickiewicza 30, 30-059 Krakow, Poland\\

vladimir@mat.agh.edu.pl    }

\end{center}

\vspace{3mm}

 \noindent{\bf Abstract.}
We present  new  analytical solutions to  the hyperbolic
generalization of Burgers equation, describing interaction of the
wave fronts. To obtain them, we employ a modified version of the
Hirota method.

 \vspace{3mm}


\vspace{3mm}


\section{ Introduction }

At the present time there are known only few analytical methods
enabling to solve sufficiently general sets of initial or boundary
value problems for nonlinear PDEs and they are applied to either
completely integrable  PDEs \cite{Dodd} or those PDEs that can be
transformed to linear ones by means of non-local change of
variables \cite{nonloclin}.

The majority of evolutionary PDEs used to simulate non-linear
transport phenomena is not completely integrable. Yet it is well
known that under certain conditions coherent structures formation
take place during the transport phenomena  occurring in open
dissipative systems \cite{opendis} and their analytical
description  is of great interest.

The simplest coherent structures are represented by periodic,
quasiperoidic, kink-like and soliton-like travelling wave (TW)
solutions. In recent decades a  number of effective
methods, enabling to obtain analytical expressions for  TW
solutions describing coherent structures have been put forward
\cite{Fushchich1, Fushchich2,
Clarkson, ClarksMans, OlvRos, Vorobjev, Fan,BarYur, NikBar,
Vladku1}. But most of the papers dealing with this subject
concentrate upon the finding out solutions, evolving in a
self-similar mode. And only few authors  look for
"bi-soliton" (or  a "multi-soliton") solutions to  those PDEs,
that are not completely integrable. Being informal, we
mean by this term solutions describing interaction of TW regimes.

Perhaps, the most advanced study of   bi-soliton solutions of PDEs
that are not completely integrable has been performed by
H.~Cornille and A.~Gervois \cite{Cornille1,Cornille2}. They
described a broad class of non-linear PDEs, possessing bi-soliton
solutions. Yet the methodology put forward by these authors
applies merely to so called "factorized" PDEs and, besides,
further investigations revealed that their classification  is not
complete. T.~~Kawahara and M.~~Tanaka considered equation of the Fisher type
that does not fit the classification scheme from
\cite{Cornille2}. They succeeded in analytical description of
travelling fronts, using  the classical Hirota method
\cite{Kawahara}.

In this work we present the analytical description of interacting
wave fronts within the hyperbolic generalization of Burgers
equation, using the modified Hirota method.  The structure of the
study is following. In section~2 we introduce a canonical form of
generalized Burgers equation and study its TW solutions within the
classical Hirota method. At the beginning of
section~3, using the preliminary information about the TW regimes,
we try to obtain bi-soliton solutions using the classical Hirota
method [Dodd] and show that it is rather impossible. So in the following
part of this section we use  a modified version of
the Hirota method, enabling to achieve our goal.

\section{Hyperbolic generalization of Burgers equation and its TW solutions}

We consider the hyperbolic  generalization of  Burgers equation
 \cite{Makar}:
\begin{equation}\label{GBEaux}
\tau u_{tt}+u_t+u\,u_x-\kappa\,u_{xx}=\sum_{\nu=0}^3a_\nu\,u^\nu.
\end{equation}
Here $u=u(t,\,x)$, $\tau,\,\,\kappa,\,\, a_\nu$ are constant
parameters (for physical reason $\,\tau\geq\,0 $, and
$\kappa\,>\,0$). Lower indices  denote partial derivatives with
respect to corresponding variables. Assuming that the polynomial
in RHS of equation (\ref{GBEaux}) has three real roots, we can
rewrite this equation in the following equivalent form:
\begin{equation}\label{GBE}
\tau u_{tt}+u_t+u\,u_x+B\,u_x-\kappa\,u_{xx}=\lambda\,u\,\left(u-S
\right)\,\left(u-Q \right),
\end{equation}
where $B,\,S,\,Q,\,\lambda$ are constant parameters. We'll use for
equation (\ref{GBE}) the abbreviation GBE.

As it was previously announced, the main goal of this study is to
obtain the analytical description to the bi-solitons, using the
Hirota method. Employment of this method, that will be briefly
described later on, leads to a very complicated system of
non-linear algebraic equations that rather could not be solved
directly without any additional information about parameters being
involved into the scheme. Processing of the system of algebraic
equations appearing within the Hirota method becomes more simple
if we restrict ourself to finding out solutions, possessing some
already known asymptotic features. We will assume in this study
that bi-soliton solutions tend on   $+ \infty$ or (and) $ -\infty$
to corresponding TW solutions. For this reason we begin with the
analytical description of kink-like  TW solutions, playing the
role of asymptotics. To obtain them, we use the ansatz
\begin{equation}\label{Hir_ans}
u=\frac{f_x}{f}.
\end{equation}
Following the Hirota method, we put
\[
f=1+\epsilon\,\varphi(t,\,x),
\]
where $\varphi(t,\,x)$ is an unknown function and $\epsilon$ is a
formally small parameter.  We say that $\varphi(t,\,x)$ is
"formally" small, since after the substitution of (\ref{Hir_ans})
into (\ref{GBE}) we use the asymptotic decomposition of the
expression obtained, treating this parameter as a small one, but
finally we put $\epsilon=1$. The detailed description of this
procedure can be found e.g. in \cite{Dodd}.

So, we insert ansatz (\ref{Hir_ans}) into (\ref{GBE}) and multiply
the obtained expression by $f^3.$ This way a third-order
polynomial with respect to $\epsilon$ is obtained. Equating to
zero the coefficient of $\epsilon^1$, we get the following PDE:
\begin{equation}\label{exp_sol}
\tau\,\varphi_{ttx}+\varphi_{tx}+B\,\varphi_{xx}-\kappa\,\varphi_{xxx}+Q\,S\,\lambda\,\varphi_x=0.
\end{equation}
It is obvious, that  solutions of the linear PDE (\ref{exp_sol})
can be presented in the form
\begin{equation}\label{fi}
\varphi(t,\,x)=exp\left[a\,x-v\,t+c\right],
\end{equation}
where  $a,\,\,v,\,\,c$ are constant parameters. Inserting
(\ref{fi}) into the ansatz (\ref{Hir_ans}), and equating to zero
coefficients of $\epsilon^k,\,\,k=1,\,2,\,3$, we obtain the
following system of algebraic equations:
\begin{eqnarray}
-a\,B+v+a^2\,\kappa-Q\,S\,\lambda-v^2\,\tau=0, \label{eps1} \\
-v+a^2(1+\kappa)+2\,Q\,S\,\lambda+a\left[ B+\left(S-Q
\right)\lambda\right]-v^2\tau=0, \label{eps2} \\
\left(a+Q  \right)\lambda\left(a-S  \right)=0. \label{eps3}
\end{eqnarray}
We see from equation (\ref{eps3}) that in case when $\lambda$ is
non-zero parameter $a$ is equal to either $S$ or $-Q$.

In the first case we get the solution of the form
\begin{eqnarray}
\varphi_1(t,\,x)=exp\left[S\,x-t\,v  +c_1\right], \label{fi1} \\
\nonumber \\ v=\frac{S\left[2\,B+S\left(\lambda+1
\right)+2\,\lambda\,Q\right]}{2}. \label{v}
\end{eqnarray} This function being inserted into (\ref{Hir_ans})
gives rise to a kink-like solution in case when an extra condition
\begin{equation}\label{kapv}
\kappa=\kappa_v=\frac{\tau\left[2\,B+S+\lambda\left(2\,Q+S\right)\right]^2-2\left(1+\lambda\right)}{4}
\end{equation}
is satisfied. As it was mentioned before, we finally put
$\epsilon=1$ in the formula (\ref{Hir_ans}). This may be done
correctly because the parameter $c$ in (\ref{fi}) is arbitrary. So
we can "rescale" it as follows: $c\rightarrow c+log\,1/\epsilon$.
This trick eliminates $\epsilon$ from the formula (\ref{Hir_ans}).

In the second case, i.e. when $a=-Q$ we get the following solution
of system (\ref{eps1})--(\ref{eps2}):
\begin{eqnarray}
\varphi_2(t,\,x)=exp\left[-Q\,x-t\,w  +c_2\right], \label{fi2} \\
\nonumber \\ w=\frac{-Q\left[2\,B-Q\left(\lambda+1
\right)-2\,\lambda\,S\right]}{2} \label{w},
\end{eqnarray}
which leads to the kink-like TW solution providing that
\begin{equation}\label{kapw}
\kappa=\kappa_w=\frac{\tau\left[\lambda\left(2\,S+Q\right)+Q-2\,B\right]^2-2\left(1+\lambda\right)}{4}.
\end{equation}
We see that, generally speaking, parameters $\kappa_v$ and
$\kappa_w$ are different. But, in contrast to $a,\,v,\,w\quad
\mbox{and}\quad c_i,\,\,i=1,\,2,$ the parameters $\kappa$
characterizes equation (\ref{GBE}) and cannot be chosen arbitrarily. So  in
order that (\ref{fi1}) and (\ref{fi2}) describe two different
solutions of GBE, we should equalize $ \kappa_v$ and $\kappa_w$. It is
easy to show that equalities
\begin{equation}\label{commonkap}
\kappa_v=\kappa_w=\kappa=\frac{\left(Q+S \right)^2 \tau
\left(1+3\,\lambda\right)-
  8\left(1+\lambda \right)}{16}
  \end{equation}
take place if
\begin{equation}\label{parB}
B=\frac{\left(S-Q \right)\left(\lambda-1 \right)}{4}.
\end{equation}

So when $\kappa$ and $B$ are given by equations (\ref{commonkap})
and (\ref{parB}) correspondingly, we get two different kink-like
TW solutions of the same equation.

\section{Bi-soliton solutions}

\subsection{Procedure based on the classical Hirota method }

We look for the solution of the following form:
\begin{equation}\label{bisolgen}
u(t,\,x)=F\left(\omega_1,\,\omega_2   \right),
\end{equation}
where $ \omega_1=S\,x-v\,t+c_1$, $\omega_2=-Q\,x-w\,t+c_2$.
Solution (\ref{bisolgen}) describes interactions of various TW in
cases when $\omega_1$ and $\omega_2$ are not proportional. As it
was mention before, in our attempts to obtain bi-soliton solutions
within the Hirota method, we face the necessity to solve very
difficult systems of non-linear algebraic equations. To make the
problem more tractable, we should pose some conditions on the
parameters to be determined. So assuming that solution  we are
looking for describes asymptotically (depending on the sort of
interaction) one or two TW fronts, we choose parameters $v$ and
$w$ in accordance with the formulae (\ref{v}) and (\ref{w}). Of
course, we can proceed this way further on and incorporate the
formulae (\ref{commonkap}), (\ref{parB}) as well, but our analysis
shows that in this case there is a lack of non-trivial solutions.

When looking for  simplest bi-soliton solution,  it is
instructive to choose $f$ in the form of superposition of
functions corresponding to the TW solutions:
\begin{equation}\label{clas2solf}
f+\epsilon\left(
exp[\omega_1]+exp[\omega_2]\right)+R\,\epsilon^2\,
exp[\omega_1+\omega_2].
\end{equation}
Inserting (\ref{Hir_ans}) with $f$ given by the formula
(\ref{clas2solf}) into equation (\ref{GBE}) and multiplying the
obtained expression by $f^3,$ we get a six-order polynomial with
respect to $\epsilon$. Equating to zero coefficients of
$\epsilon^n,\,\,n=1,\,2,...6,$ we obtain six nonlinear algebraic
equations (we denote them as (eq1),...(eq6)), containing the
parameters and  products of $exp[\omega_1]$ and $exp[\omega_2]$.
For brevity we shall use the notation
\[
x^m=exp[m\cdot\omega_1],\,\,\,\,y^n=exp[n\cdot\omega_2].
\]

Calculations were carried out in several steps essentially based
upon employment of the  "Mathematica" software. Below we describe
a crucial points of  the procedure employed, enabling to repeate
it. Yet it is obvious, that correctness of solutions obtained within
this procedure can be checked by direct inspection without going
into details.

Before we proceed further on, let us note that we do not take into
account solutions with $Q=0,$ $S=0,$ $Q=-S$ and $\lambda=-1/3$.
Elementary analysis of formula (\ref{Hir_ans}) whth (\ref{v}) and
(\ref{w}) being taken into account shows, that
in all these cases we get at most TW solution.

So let us describe the procedure of obtaining bi-soliton solution
based on the ansatz (\ref{Hir_ans}).

\begin{itemize}
\item[\bf{Step 1.}] First of all we consider  (eq6), which is the
simplest one:
\begin{equation}\label{eq6-1}
Q\,R^3\,S\,\left(S-Q \right)x^3\,y^3\,\lambda=0.
\end{equation}
From this equation we obtain that
\begin{equation}\label{step1a}
S=Q.
\end{equation}

\item[\bf{Step 2.}] Next we consider (eq1), which is a linear one
with respect to variables $x,\,\,y.$ Taking  into account
(\ref{step1a}) and equating to zero coefficients of $x$ and $y$,
we obtain the following system:
\begin{eqnarray}
-2\left(1+\lambda \right)-4\,\kappa+\tau\left[4\,B^2+
4\,B\,S\left(1+3\,\lambda\right)+S^2\left(1+6\,\lambda+9\,\lambda^2 \right)\right]=0, \label{step2a} \\
2\,\left(1+\lambda\right)+4\,\kappa+\tau\,\left[
-4B^2+4\,B\,S\left(1+3\,\lambda\right)-S^2\left(1+6\,\lambda+9\,\lambda^2
\right)\right]=0.
 \label{step2b}
\end{eqnarray}
Summing up these two equations we get the equation
\begin{equation}\label{Bzero}
8\,B\,S\,\tau\,\left(1+3\, \lambda\right)=0.
\end{equation}
To fulfill (\ref{Bzero}), we put  $B=0$. Returning then to
equation (\ref{step2a}) we obtain the following expression for
$\kappa$:
\[
\kappa=\frac{-2\left(1+\lambda
\right)+\tau\,S^2\left(1+6\,\lambda+9\,\lambda^2 \right)}{4}.
\]

\item[\bf{Step 3.}] With these conditions  (eq1), (eq2) and (eq6)
nullify while subsystem (eq3)--(eq5) becomes as follows:
\begin{eqnarray}
S^3\,x\,y\left(x-y \right)\left(1+3\,\lambda\right)
\left[2\,S^2\,\tau\left(1-R\right)\left(1+3\,\lambda \right)+R
\right]=0,
\label{step3a} \\
R\,S^3\,x\,y\left(x^2-y^2 \right)\left(1+3\,\lambda\right)=0, \label{step3b} \\
R^2\,S^3\,x^2\,y^2\left(x-y \right)\left(1+3\,\lambda\right)=0.
\label{step3c}
\end{eqnarray}
\end{itemize}

\begin{figure}
\includegraphics[width=3 in, height=2 in]{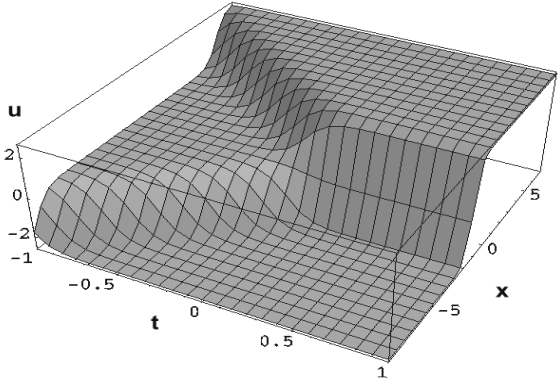}
\caption{An example of a bi-soliton solution of  equation (\ref{BE}),
described by the formula (\ref{bisolBE}) for
$S=2.5,\,c_1=2,\,c_2=-6,\,\kappa=0.5$}\label{Fig:1}
\end{figure}

The only non-trivial  solution to (\ref{step3a})-(\ref{step3c})
corresponding to bi-solitons is $R=\tau=0$.

It can be easily checked starting from equation (\ref{Bzero}) that
we do not obtain any other solution choosing solution $\tau=0$
instead of $B=0$ in the second step.

Thus, we became convinced that there is impossible to obtain a
bi-soliton solution satisfying (\ref{GBE}), using the classical
Hirota method. Yet as a by-product we get the following result:

\vspace{3mm}

{\bf Theorem 1. }
{\it The  Burgers equation
\begin{equation}\label{BE}
u_t+u\,u_x-\kappa\,u_{xx}=-\left(1+2\,\kappa
\right)\,u\,\left(u^2-S^2 \right)
\end{equation}
possess a bi-soliton solution
\begin{equation}\label{bisolBE}
u(t,\,x)=S\,\frac{exp\left( S\,x-v\,t +c_1\right)-exp\left(
-S\,x-v\,t +c_2\right)}{1+exp\left( S\,x-v\,t
+c_1\right)+exp\left( -S\,x-v\,t+c_2\right)}
\end{equation}
where $v=-S^2\,(1+3\,\kappa)$.
}

\vspace{3mm}

Example of bi-soliton solution describing interaction of two wave
fronts is shown in Fig.~~\ref{Fig:1}.

\subsection{Procedure based on the modified Hirota method }

One can succeed in obtaining bi-soliton solutions to (\ref{GBE})
by slightly modifying Hirota method. Let us consider the following
ansatz:
\begin{equation}\label{Hir_modif}
u(t,\,x)=\frac{g}{f},
\end{equation}
in which we put
\begin{eqnarray}
g=\epsilon\left[\alpha\,exp\left(\omega_1
\right)+\beta\,exp\left(\omega_2
\right)\right]+\epsilon^2\,A\,exp\left(\omega_1+\omega_2 \right),
\label{g}\\\nonumber\\ f=1+\epsilon\left[exp\left(\omega_1
\right)+exp\left(\omega_2
\right)\right]+\epsilon^2\,R\,exp\left(\omega_1+\omega_2 \right) \label{f}.
\end{eqnarray}
So we  insert (\ref{Hir_modif}) into (\ref{GBE}), multiply
the resulting equation by $f^3$ and obtain this way a six-order
polynomial with respect to $\epsilon.$ Equating to zero
coefficients of $\epsilon^k,\,\,\,k=1,2,...6$,  we get a system of
non-linear algebraic equations which we again denote by
${(eq1)}$--${(eq6)}$.  Using the "Mathematica" software  we
succeeded in obtaining a non-trivial solustion of these equations.
A sketch of the procedure employed is presented below.

\begin{itemize}

\item[\bf{Step 1.}] We consider (eq6) which reads as follows:
\[
-A\,\left(A+Q\,R  \right) \left(A-S\,R  \right)x^3 y^3=0.
\]
In order to satisfy this equation, we put $A=R\,S$.

\item[\bf{Step 2.}]  Next we consider (eq1).
Nullifying the coefficient of $y^1$ we obtain the equation
\begin{eqnarray}
Q^2\,\beta\,\left\{  -2\left(  1+2\,\kappa \right)+\tau\left(
4\,B^2-4\,B\,Q+Q^2\right)+\left(Q+2\,S \right)^2\lambda^2\tau+
\right.
 \nonumber \\
     \left. +\lambda\left[-2\left(1-Q^2\tau\right)+4\,Q\,S\,\tau-4\,B\,\tau\left(Q+2\,S\right)\right]\right\}=0.\nonumber
\end{eqnarray}
This equation will be satisfied if we choose $\beta=0.$ With this
choice we consider (eq5) and, nullifying coefficient of $x^3\,y^2$
obtain:
\begin{eqnarray}
\left(
S-\alpha\right)\left[4\,Q\,S-4\,S^2\lambda+\tau\,Q^4\left(1+\lambda
\right)^2+4\,Q^3\tau\left(1+\lambda \right)\left(S\,\lambda-B
\right)+ \right.\nonumber \\
\left.
+2\,Q^2\left(1-2\,\kappa+\lambda+2\,B^2\tau-4\,B\,S\,\lambda\,\tau+2\,S^2\lambda^2\tau\right)
\right]=0. \nonumber
\end{eqnarray}
This equation will be satisfied if $S=\alpha.$    Equating to zero
the remaining term, we get:
\begin{eqnarray}
-\frac{1}{4}R^2S^3x^2y^3\left\{-2-4\,\kappa+4\,B^2\tau+4\,B\,S\,\tau+S^2\tau+\left(2\,Q+S\right)^2\lambda^2\tau+
\right.\label{interm1}
\\
\left.+2\,\lambda\left[-1+2\,Q\,S\,\tau+S^2\tau+2\,B\left(2\,Q+S\right)\tau\right]\right\}=0.
\nonumber
\end{eqnarray}
Equation (\ref{interm1}) is satisfied if  $\kappa$ is expressed by
the formula (\ref{kapv}).

\item[\bf{Step 3.}]   Next we consider (eq3). The coefficient of $x^2y$ reads as
follows:
\[
-\frac{1}{4}\,Q\left(R-1\right)S\left(Q+S\right)^2\left(1+3\,\lambda\right)\tau\left(4\,B+2\,S-Q+\lambda\,Q+2\,\lambda\,S\right)
\]
In order to nullify it we  put
\begin{equation}\label{exprforB}
B=-\frac{2\,S-Q+\lambda\left(Q+2\,S\right)}{4}.
\end{equation}
The remaining coefficient of (eq3) to be nullified is that of
$x\,y^2$. Taking (\ref{exprforB}) into account we can write it as
follows:
\[
-Q\left(R-1\right)S\left(Q+S+Q\,\lambda+2\,S\,\lambda\right).
\]
Analysis of formula (\ref{Hir_modif}) shows that option $R=1$
leads to a one-soliton solution so we nullify this expression
assuming that
\begin{equation}\label{exprforlam}
\lambda=-\frac{Q+S}{Q+2\,S}.
\end{equation}

\item[\bf{Step 4.}] Taking into account all the above conditions
and repeating the procedure we conclude that only (eq2) and (eq4)
are nonzero: (eq2) is given by the formula
\[
\frac{Q^2\left(R-1\right)S^2x\,y\left[2\,Q+4\,S+4\,Q^2\tau+
8\,Q^2S\,\tau+5\,Q\,S^2\tau+\tau\,S^3\right]}{2\left(Q+2\,S\right)^2},
\]
while (eq4) is as follows:
\[
\frac{R\,Q^2\left(R-1\right)S^2x^2\,y^2\left[2\,Q+4\,S+4\,Q^2\tau+
8\,Q^2S\,\tau+5\,Q\,S^2\tau+\tau\,S^3\right]}{2\left(Q+2\,S\right)^2}.
\]
Both of them are nullified if
\begin{equation}\label{exprfortau}
\tau=-\frac{2\left(Q+2\,S\right)}{\left(Q+S\right)\left(2\,Q+S\right)^2}.
\end{equation}

\end{itemize}

\begin{figure}
\includegraphics[width=3 in, height=2 in]{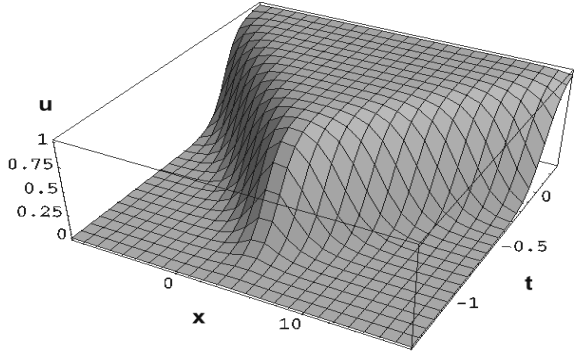}
\caption{Creation  of two wave fronts "from nothing", described by
the formula
 (\ref{bi-sol-sol}) for $S=1,\,Q=-1.9,c_1=10,\,c_2=-6,\,R=0$}\label{Fig:2}
\end{figure}

\begin{figure}
\includegraphics[width=3 in, height=2 in]{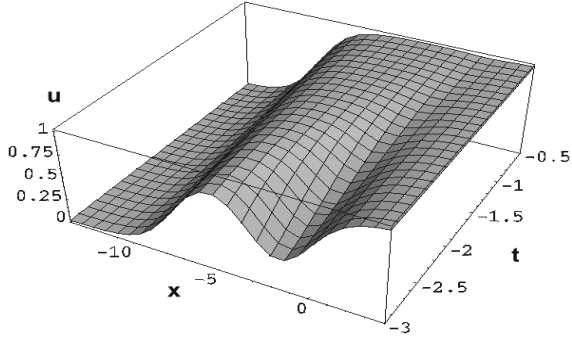}
\caption{Interaction of localized TW solution and kink described
by the formula (\ref{bi-sol-sol}) for
$S=1,\,Q=2,\,c_1=2,\,c_2=-3,\,R=1500$}\label{Fig:3}
\end{figure}

And this way we obtained the bi-soliton solution to GBE.
Elementary but slightly tiresome analysis shows that system (GBE)
does not admit any other bi-soliton solution described by the
formula (\ref{Hir_modif}). The only  exception is the solution
connected with the already obtained  solution by means of the
transformation
$S\,\mapsto\,Q,\,\, Q\,\mapsto\,-S,\,\, \alpha=0, \,\,\beta=-Q.$
But existence of the second solution is a mere consequence of the
arbitrariness of the choice of what is $\omega_1$ and what is
$\omega_2$ in the formulae (\ref{g})--(\ref{f}). It is obvious then,
that this way we don't get any other independent solution.

Let us summarize the result obtained as the following statement.

\vspace{3mm}

{\bf Theorem 2.}
{\it If
\[
B=\frac{2\,Q-S}{4},\quad \lambda=-\frac{Q+S}{Q+2\,S},\quad
\tau=-\frac{2\,(Q+2\,S)}{(Q+S)(2\,Q+S)^2}
\]
and
\[
\kappa=-\frac{Q+2\,S}{8\,(Q+S)}
\]
then equation (\ref{GBE}) admits the following bi-soliton solution
\begin{equation}\label{bi-sol-sol}
u(t,\,x)= \frac{S\,
exp\left(\omega_1\right)\left[1+R\,exp\left(\omega_2\right)\right]}
{1+exp\left(\omega_1\right)+exp\left(\omega_2\right)+R\,exp\left(\omega_1+\omega_2\right)},
\end{equation}
where
\[
\omega_1=S\left[x+\frac{t}{4}\,\frac{Q\left(2\,Q+S\right)}{Q+2\,S}\right]+c_1,
\quad
\omega_2=-Q\left[x-\frac{t}{4}\,\left(2\,Q+S\right)\right]+c_2,
\]
$S,\,\,Q,\,\,R,\,\,c_1,\,\,c_2$ are arbitrary constants.
}

\vspace{3mm}


Patterns of the interacting wave fronts described by the formula
(\ref{bi-sol-sol}) are shown in Figs.~~\ref{Fig:2},~~\ref{Fig:3}.

\section{Concluding Remarks}

Thus, we have obtained bi-soliton solutions for classical Burgers
equation with cubic non-linearity and for the GBE. To our
knowledge, such solutions  for the equation (\ref{GBE}) have been
obtained for the first time. In this study we looked for the
simplest bi-soliton solutions described by the formula
(\ref{Hir_modif}). It is obvious that rational solutions of this
sort can be unlimitedly generalized. However it  should be noted,
that one faces very non-trivial technical problems when
incorporating additional terms into (\ref{g})--(\ref{f}).  And for
successful application of
the above method it is desired to have some extra
information about the possible values of the parameters.
Beside the asymptotic analysis, one can incorporate another
methods, such as the Painlev\'{e} analysis \cite{Tabor} and study
of conserved quantities \cite{Bluman}.

Let us mention in conclusion that solutions obtained in this work
describe  merely interaction of kinks. It would be interesting to
obtain an analytical description of interacting  TW waves of
another sort. In paper \cite{Vladku2} there were obtained
periodic, soliton-like and some other solutions to GBE. All of
them  have been obtained within the methods different from that
used in this paper. Consequently a successful investigations of
the interactions of TW  of different types should be based on
employment of  more sophisticated ansatzes.

\end{document}